\documentclass[10pt,letterpaper]{article}
\usepackage{opex3_liheng}
\usepackage{color}

\usepackage{bbm}
\usepackage{mathrsfs}
\usepackage{amsfonts}
\usepackage{amssymb}
\usepackage{graphicx}
\usepackage{cite}
\usepackage[linesnumbered,ruled,lined]{algorithm2e}
\usepackage{array}
\usepackage{subfigure}
\usepackage{multirow}
\usepackage{amsmath}
\usepackage{bm}
\usepackage{upgreek}
\usepackage{float}
\newcommand{\format}[1]{\text{\bf #1}}

\usepackage{hyperref}

\begin{document}

\title{Motion-corrected Fourier ptychography}

\author{Liheng Bian$^1$, Guoan Zheng$^2$, Kaikai Guo$^2$, Jinli Suo$^{1,*}$, Changhuei Yang$^3$, Feng Chen$^1$ and Qionghai Dai$^1$}

\address{$^1$Department of Automation, Tsinghua University, Beijing 100084, China\\
$^2$Department of Biomedical Engineering, University of Connecticut, Storrs, Connecticut 06269, USA\\
$^3$Department of Electrical Engineering, California Institute of Technology, Pasadena, CA 91125, USA}

\email{$^*$jlsuo@tsinghua.edu.cn} 

\homepage{https://sites.google.com/site/lihengbian/} 


\begin{abstract}
Fourier ptychography (FP) is a recently proposed computational imaging technique for high space-bandwidth product imaging. In real setups such as endoscope and transmission electron microscope, the common sample motion largely degrades the FP reconstruction and limits its practicability. In this paper, we propose a novel FP reconstruction method to efficiently correct for unknown sample motion. Specifically, we adaptively update the sample's Fourier spectrum from low spatial-frequency regions towards high spatial-frequency ones, with an additional motion recovery and phase-offset compensation procedure for each sub-spectrum. Benefiting from the phase retrieval redundancy theory, the required large overlap between adjacent sub-spectra offers an accurate guide for successful motion recovery. Experimental results on both simulated data and real captured data show that the proposed method can correct for unknown sample motion with its standard deviation being up to 10$\%$ of the field-of-view scale. We have released our source code for non-commercial use, and it may find wide applications in related FP platforms such as endoscopy and transmission electron microscopy.
\end{abstract}


\bibliographystyle{osajnl_suo}
\bibliography{mcFP}

%
%

\section{Introduction}\label{sec:Introduction}

\begin{figure}[t]
\centering
\centerline{\includegraphics[width=1.3\textwidth]{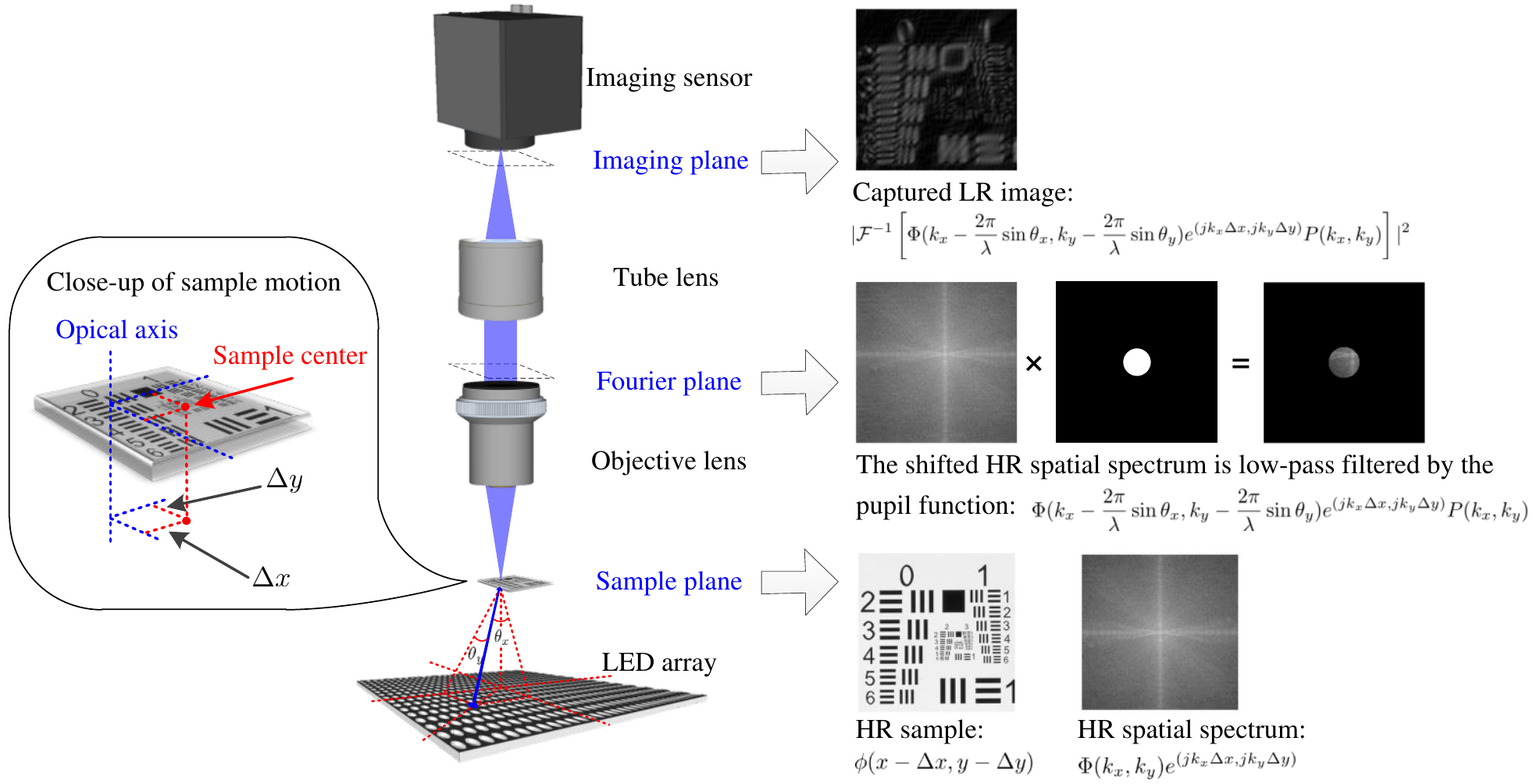}}
\caption{The FPM system and its image formation in the case with sample motion.}
\label{fig:Fig_System}
\end{figure}

Fourier ptychography (FP) is a novel computational imaging technique for high space-bandwidth product (SBP) imaging \cite{FPM_Nature, FPM_Quantitative}. This technique captures a set of low-resolution (LR) images, which correspond to different Fourier sub-spectra of the sample. By stitching these sub-spectra together in Fourier space using a reconstruction algorithm, a large field-of-view (FOV) and high-resolution (HR) image of the sample can be obtained. It has been successfully demonstrated in optical microscopy as Fourier ptychographic microscopy (FPM) \cite{FPM_Nature}, where the incident light is assumed to be a plane wave, and the LR images are captured under different incident angles from the LEDs placed at different locations. The measurements correspond to different Fourier sub-spectra of the sample, as shown in Fig. \ref{fig:Fig_System}. The synthetic numerical aperture (NA) of the FPM setup reported in Ref. \cite{FPM_Nature} is $\sim$0.5, and the FOV reaches $\sim$120 mm$^2$, which greatly improve the throughput of existing microscope. Due to its simple setup and super performance, FPM has been widely applied in 3D imaging \cite{3D_1, 3D_2}, fluorescence imaging \cite{Fluo_1, Fluo_2}, mobile microscope \cite{Cellphone_1, Cellphone_2}, and high-speed in vitro imaging \cite{Vitro, LaserFPM}.

Fourier ptychographic reconstruction is a typical phase retrieval optimization process, which needs to recover a complex function given the intensity measurements of its linear transforms (Fourier transform in FP) \cite{Phase_Review}. Existing FP algorithms can be classified into three categories including the alternating projection (AP) method \cite{FPM_Nature, FPM_Quantitative, PupilFunction}, the semi-definite programming (SDP) based method \cite{FPMAlgorithm_Convex}, and the non-convex gradient descent method \cite{FPMAlgorithm_Bian, FPMAlgorithm_Laura, TPWFP}. AP adds constraints alternately in spatial space (captured intensity images) and Fourier space (pupil function), to stitch the LR sub-spectra together. Ou et al. \cite{PupilFunction} added an additional pupil function update procedure into AP to correct pupil function error. AP is easy to implement and fast to converge, but is sensitive to measurement noise and system errors, which arise from numerous factors such as short camera exposure time \cite{FPMAlgorithm_Bian} and LED misalignment. To tackle measurement noise, Bian et al. \cite{FPMAlgorithm_Bian} proposed the Wirtinger flow optimization for Fourier ptychography (WFP), which uses the gradient-descent scheme and the Wirtinger calculus \cite{Phase_Wirtinger} to minimize the intensity errors between estimated LR images and measurements. WFP is robust to Gaussian noise, and can produce better reconstruction results in low-exposure imaging scenarios. Thus it can largely decrease the required image acquisition time. However, it needs careful initialization since the optimization is non-convex. Later, Yeh et al. \cite{FPMAlgorithm_Laura} tested different objective functions (intensity based, amplitude based and Poisson maximum likelihood) under the gradient-descent optimization scheme for FP reconstruction. The results show that the amplitude-based and Poisson maximum-likelihood objective functions produce better results than the intensity-based objective function. To address the LED misalignment, the authors added a simulated annealing optimization procedure into each iteration to search for the optimal pupil locations. Recently, Bian et al. \cite{TPWFP} proposed a novel FP reconstruction method termed truncated Poisson Wirtinger Fourier ptychographic reconstruction (TPWFP), which utilizes Poisson maximum likelihood for better signal modeling, and truncated Wirtinger gradient for effective error removal. The method can efficiently handle pupil location error and various measurement noise including Poisson noise, Gaussian noise and speckle noise. To make the FP reconstruction convex, Horstmeyer et al. \cite{FPMAlgorithm_Convex} proposed an SDP \cite{Phase_Lift_2, Phase_Cut} based optimization method to recover the HR spatial spectrum. The method is robust to Gaussian noise and guarantees a global optimum, but converges slowly which makes it impractical in real applications.

The methods discussed above tackle different challenges in FP reconstruction. However, none of them is able to correct for common sample motion during the multiple acquisition process. For endoscopy applications, hand-held endoscope probes may move during multiple acquisitions of the same sample \cite{swift2005method, vercauteren2006robust}. In transmission electron microscopy (TEM), sample drift is a common problem for multiple image acquisitions \cite{li2013electron, liao2013structure}. Therefore, tackling the sample-motion problem is a crucial step for applying the FP scheme to high-resolution endoscopy \cite{pacheco2016reflective, kaikai2016fourier} and TEM (aperture scanning FP \cite{3D_1}). 
Also, as shown in the following experiments, sample motion degrades the conventional reconstruction a lot. 
To tackle this problem, we propose a novel FP reconstruction method in this paper, termed motion-corrected Fourier ptychography (mcFP). This technique adaptively updates the HR spatial spectrum loop by loop from low spatial-frequency regions towards high spatial-frequency ones, similar to Ref. \cite{AFP}. Required by the phase retrieval redundancy theory that measurements should be much more than the to-be-recovered signals \cite{dong2014sparsely, Phase_Wirtinger, FPMAlgorithm_Bian}, there is a large overlap between adjacent sub-spectra in the last and the current loop. 
Benefiting from this overlap region which is already updated in the last loop, for each sub-spectrum to be updated in the current loop, we can successfully search for its unknown motion shift by minimizing the difference between the captured image and corresponding reconstruction. Then the obtained motion shift is utilized to compensate the phase offset of the sub-spectrum. Until all the sub-spectra are updated, the current iteration is completed. After several such iterations (2 iterations are enough for convergence as the experimental results indicate), we can recover the high-quality HR spatial spectrum without sample-motion degeneration.
To validate the effectiveness of mcFP, we test it on both simulated data and real captured data. Both experimental results show that mcFP can correct for sample motion with its standard deviation being up to 10$\%$ of the field-of-view scale. As far as we know, this is the first study that addresses sample motion in FP. It may provide new insights and find important applications in various Fourier ptychographic imaging platforms such as endoscopy and TEM.

\section{Methods}

In the following , we use FPM as an example to demonstrate the mathematical model of the proposed mcFP method for better understanding. We first derive the image formation of FPM in the case with sample motion, and then introduce the proposed mcFP in detail.

\subsection{Image formation of FPM}

FPM is a coherent imaging system. For a relatively thin sample \cite{HighNA_Xiaoze}, its different spatial spectrum regions can be accessed by angularly varying coherent illumination. Specifically, under the assumption that the incident light is a plane wave, we can describe the light field transmitted from the sample as $\phi(x, y) e^{(jx\frac{2\pi}{\lambda}\sin\theta_x, jy\frac{2\pi}{\lambda}\sin\theta_y)}$ with no sample motion, where $\phi$ is the sample's complex spatial map, $(x, y)$ are the 2D spatial coordinates, $j$ is the imaginary unit, $\lambda$ is the wavelength of illumination, and $\theta_x$ and $\theta_y$ are the incident angles, as shown in Fig. \ref{fig:Fig_System}. 
Then the light field is Fourier transformed to the pupil plane when it travels through the objective lens, and subsequently low-pass filtered by the pupil. This process can be represented as $P(k_x, k_y)\mathcal{F}(\phi(x, y) e^{(jx\frac{2\pi}{\lambda}\sin\theta_x, jy\frac{2\pi}{\lambda}\sin\theta_y)})$, where $P(k_x, k_y)$ is the pupil function for low-pass filtering, $(k_x, k_y)$ are the 2D spatial frequency coordinates in the pupil plane, and $\mathcal{F}$ is the Fourier transform operator. Afterwards, the light field is Fourier transformed again when it passes through the tube lens to the imaging sensor. Since real imaging sensors can only capture light's intensity, the image formation of FPM with no sample motion follows
\begin{eqnarray}\label{eqs:Formation_0}
\format I &=& |\mathcal{F}^{-1}\left[P(k_x, k_y)\mathcal{F}(\phi(x, y) e^{(jx\frac{2\pi}{\lambda}\sin\theta_x, jy\frac{2\pi}{\lambda}\sin\theta_y)})\right]|^2\\ \nonumber
&=& |\mathcal{F}^{-1}\left[P(k_x, k_y)\Phi(k_x - \frac{2\pi}{\lambda}\sin\theta_x, k_y - \frac{2\pi}{\lambda}\sin\theta_y) \right]|^2,
\end{eqnarray}
where $\format I$ is the captured intensity image, $\mathcal{F}^{-1}$ is the inverse Fourier transform operator, and $\Phi$ is the spatial spectrum of the sample.

In the case with sample motion, which is denoted as $(\Delta x, \Delta y)$ in the sample plane, the light field transmitted from the sample becomes $\phi(x - \Delta x, y - \Delta y) e^{(jx\frac{2\pi}{\lambda}\sin\theta_x, jy\frac{2\pi}{\lambda}\sin\theta_y)}$. Assuming that the sample motion of the field of view (FOV) is ideally circular which means that the sample region moving outside of the FOV is the same as that moving inside, the spatial spectrum of the shifted sample correspondingly becomes $\Phi(k_x - \frac{2\pi}{\lambda}\sin\theta_x, k_y - \frac{2\pi}{\lambda}\sin\theta_y)e^{(jk_x\Delta x, jk_y\Delta y)}$ according to the shift property of Fourier transform, and the captured image is rewritten as
\begin{eqnarray}\label{eqs:Formation_1}
\format I_{motion} &=& |\mathcal{F}^{-1}\left[P(k_x, k_y)\mathcal{F}(\phi(x - \Delta x, y - \Delta  y) e^{(jx\frac{2\pi}{\lambda}\sin\theta_x, jy\frac{2\pi}{\lambda}\sin\theta_y)})\right]|^2\\ \nonumber
&=& |\mathcal{F}^{-1}\left[P(k_x, k_y)\Phi(k_x - \frac{2\pi}{\lambda}\sin\theta_x, k_y - \frac{2\pi}{\lambda}\sin\theta_y)e^{(jk_x\Delta x, jk_y\Delta y)} \right]|^2.
\end{eqnarray}
Visual explanation of the above image formation process is diagrammed in Fig. \ref{fig:Fig_System}. From Eqs. (\ref{eqs:Formation_1}) we can see that the sample motion introduces a frequency-dependent phase offset to the sample's spatial spectrum. The proposed mcFP method targets to compensate the phase offset and eliminate its negative influence on the final reconstruction.

\subsection{Motion-corrected Fourier ptychographic reconstruction (mcFP)}

\begin{figure}[t]
\centering
\centerline{\includegraphics[width=0.8\textwidth]{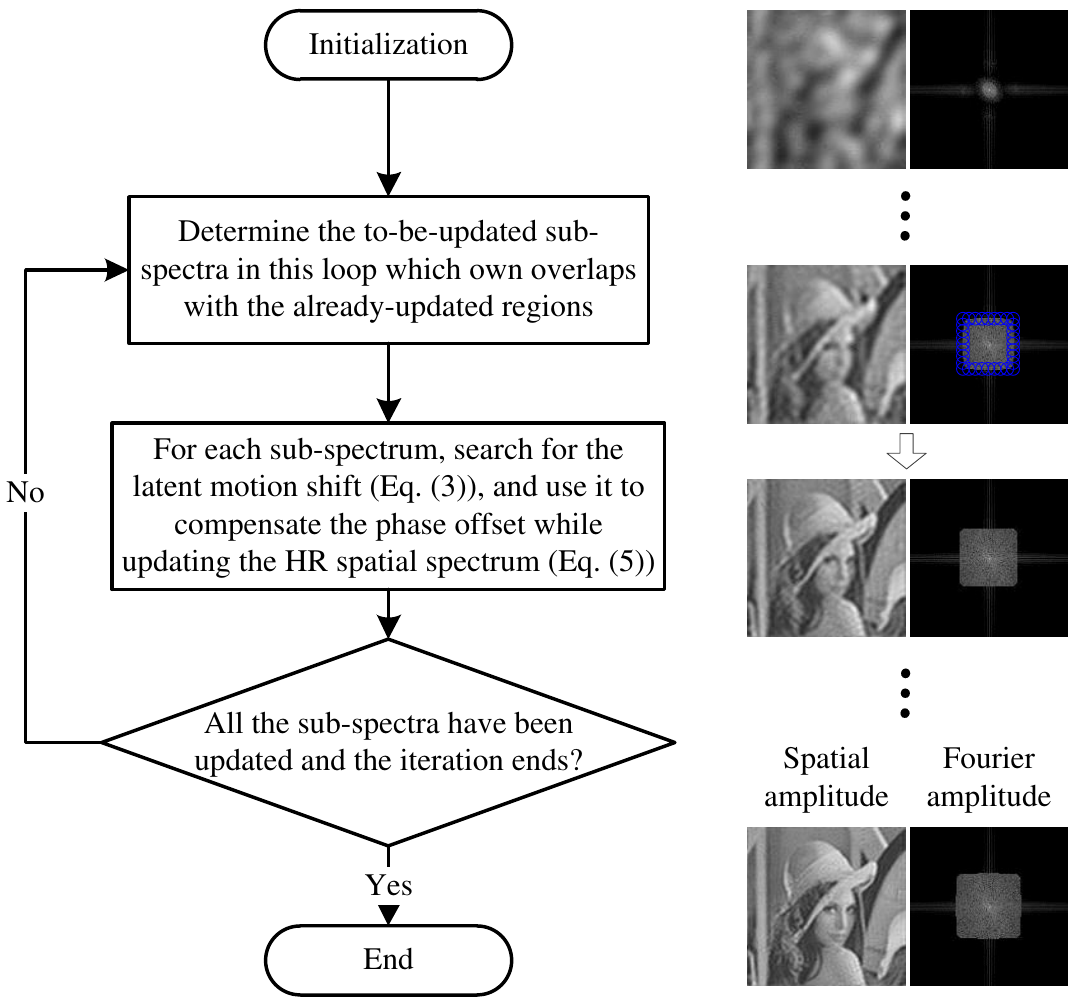}}
\caption{The algorithmic scheme of the proposed mcFP method.}
\label{fig:Fig_Algorithm}
\end{figure}

In this subsection, we begin to introduce the proposed motion-corrected Fourier ptychographic reconstruction technique in detail. The algorithm's  flow chart is diagrammed in Fig. \ref{fig:Fig_Algorithm}. Overall, we adaptively update the HR spatial spectrum in an iterative manner similar to Ref. \cite{AFP}. Each iteration includes several loops of motion recovery and spectrum updating from low spatial-frequency regions to high spatial-frequency regions.

For initialization, similar to Ref. \cite{AFP, FPMAlgorithm_Bian, TPWFP}, we use the up-sampled version of the LR image captured under normal incident light as an initial guess of the sample.
This is necessary because the LR image contains important information of the sample in the low spatial frequency region, which offers a benchmark of the sample's basic structure and guarantees the precision of the following motion recovery procedure for each to-be-updated sub-spectrum.

Then we move on to the loops of spectrum updating. At the beginning of each loop, we extend the updated sub-spectra of the last loop towards the high-frequency direction, with a fixed step size (determined by the user-defined spectrum overlapping ratio), to determine the sub-spectra to be updated in this loop. As shown on the right side in Fig. \ref{fig:Fig_Algorithm}, the blue circles in the second-row spatial spectrum indicates the determined to-be-updated sub-spectra of the current loop. Then, for each sub-spectrum $P(k_x, k_y)\Phi(k_x - \frac{2\pi}{\lambda}\sin\theta_x, k_y - \frac{2\pi}{\lambda}\sin\theta_y)$, we search for the latent motion shift $(\Delta x_{opt}, \Delta y_{opt})$ by minimizing the difference between the captured image $\format I_{motion}$ and corresponding reconstruction $|\mathcal{F}^{-1}\left[P(k_x, k_y)\Phi(k_x - \frac{2\pi}{\lambda}\sin\theta_x, k_y - \frac{2\pi}{\lambda}\sin\theta_y)e^{(jk_x\Delta x, jk_y\Delta y)} \right]|^2$. Mathematically, the searching procedure can be described as
\begin{eqnarray}\label{eqs:Search}
(\Delta x_{opt}, \Delta y_{opt}) = \mathop {\arg \min }\limits_{(\Delta x, \Delta y)} ~~~\format I_{motion} - |\mathcal{F}^{-1}\left[P(k_x, k_y)\Phi(k_x - \frac{2\pi}{\lambda}\sin\theta_x, k_y - \frac{2\pi}{\lambda}\sin\theta_y)e^{(jk_x\Delta x, jk_y\Delta y)} \right]|^2.
\end{eqnarray}
The obtained motion shift guess is utilized to compensate the phase offset of the HR spatial spectrum during its updating. The updating procedure is similar to that in Ref. \cite{PupilFunction}. Specifically, we use the square-root of the captured image to replace the amplitude of corresponding reconstruction as
\begin{eqnarray}\label{eqs:Replace}
\phi' = \sqrt{\format I_{motion}} \frac{\mathcal{F}^{-1}\left[P(k_x, k_y)\Phi(k_x - \frac{2\pi}{\lambda}\sin\theta_x, k_y - \frac{2\pi}{\lambda}\sin\theta_y)e^{(jk_x\Delta x_{opt}, jk_y\Delta y_{opt})} \right]}{|\mathcal{F}^{-1}\left[P(k_x, k_y)\Phi(k_x - \frac{2\pi}{\lambda}\sin\theta_x, k_y - \frac{2\pi}{\lambda}\sin\theta_y)e^{(jk_x\Delta x_{opt}, jk_y\Delta y_{opt})} \right]|}.
\end{eqnarray}
Then we transform $\phi'$ to Fourier space and obtain $\Phi' = \mathcal{F}(\phi')$, and use it to update the sample's HR spatial spectrum with the phase compensation $e^{(-jk_x\Delta x_{opt}, -jk_y\Delta y_{opt})}$ as 
\begin{eqnarray}\label{eqs:Update}
\Phi_{updated} = \Phi + \frac{P^*(k_x + \frac{2\pi}{\lambda}\sin\theta_x, k_y + \frac{2\pi}{\lambda}\sin\theta_y)}{|P(k_x + \frac{2\pi}{\lambda}\sin\theta_x, k_y + \frac{2\pi}{\lambda}\sin\theta_y)|_{max}^{2}}\left[\Phi'(k_x + \frac{2\pi}{\lambda}\sin\theta_x, k_y + \frac{2\pi}{\lambda}\sin\theta_y)e^{(-jk_x\Delta x_{opt}, -jk_y\Delta y_{opt})} - \Phi\right].
\end{eqnarray}
After all the sub-spectra in the current loop are updated, the loop ends and we move on to the next loop. Until all the sub-spectra corresponding to the captured images are updated, the current iteration is completed. After several such iterations, we can successfully recover the latent motion shift and the high-quality HR spatial spectrum without sample motion degeneration. Demo code of the proposed mcFP method is available at \href{http://www.sites.google.com/site/lihengbian}{\textit{http://www.sites.google.com/site/lihengbian}} for non-commercial use.


\section{Results}

In this section, we test the proposed mcFP method on both simulated and real captured data, to show its advantages over the conventional FP reconstruction using the AP algorithm.

\subsection{Quantitative metric}

To quantitatively evaluate reconstruction quality in the following simulation experiments, we utilize the relative error (RE) \cite{TWF, TPWFP} metric defined as
\begin{equation}\label{eqs:re}
RE(\textbf{z}, \hat{\textbf{z}}) = \frac{\min_{\phi \in [0, 2\pi)}||e^{-j\phi}\textbf{z} - \hat{\textbf{z}}||^2}{||\hat{\textbf{z}}||^2}.
\end{equation}
This metric describes the difference between two complex functions $\format z$ and $\hat{\textbf{z}}$. We use it here to compare the reconstructed HR complex field of the sample with its ground truth in the simulation experiments.

\subsection{Simulation experiments}\label{sec:SimulationExperiments}

In the simulation experiments, we simulate the FPM setup with its hardware parameters as follows: the NA of the objective lens is 0.08, and corresponding pupil function is an ideal binary function (all ones inside the NA circle and all zeros outside); the height from the LED plane to the sample plane is 84.8mm; the distance between adjacent LEDs is 4mm, and $15\times 15$ LEDs are used to provide a synthetic NA of $\sim$0.5; the wavelength of incident light is 625nm; and the pixel size of captured images is 0.2um. We use the 'Lena' and 'Aerial' image ($512\times512$ pixels) from the USC-SIPI image database \cite{Data} as the latent HR amplitude and phase map, respectively. The captured LR images are synthesized based on the image formation in Eqs. (\ref{eqs:Formation_1}). Their pixel numbers are set to be one eighth of the HR image along both dimensions. We repeat 20 times for each of the following simulation experiments and average their evaluations to produce final statistical results. 

\begin{figure}[t]
\centering
\centerline{\includegraphics[width=1.3\textwidth]{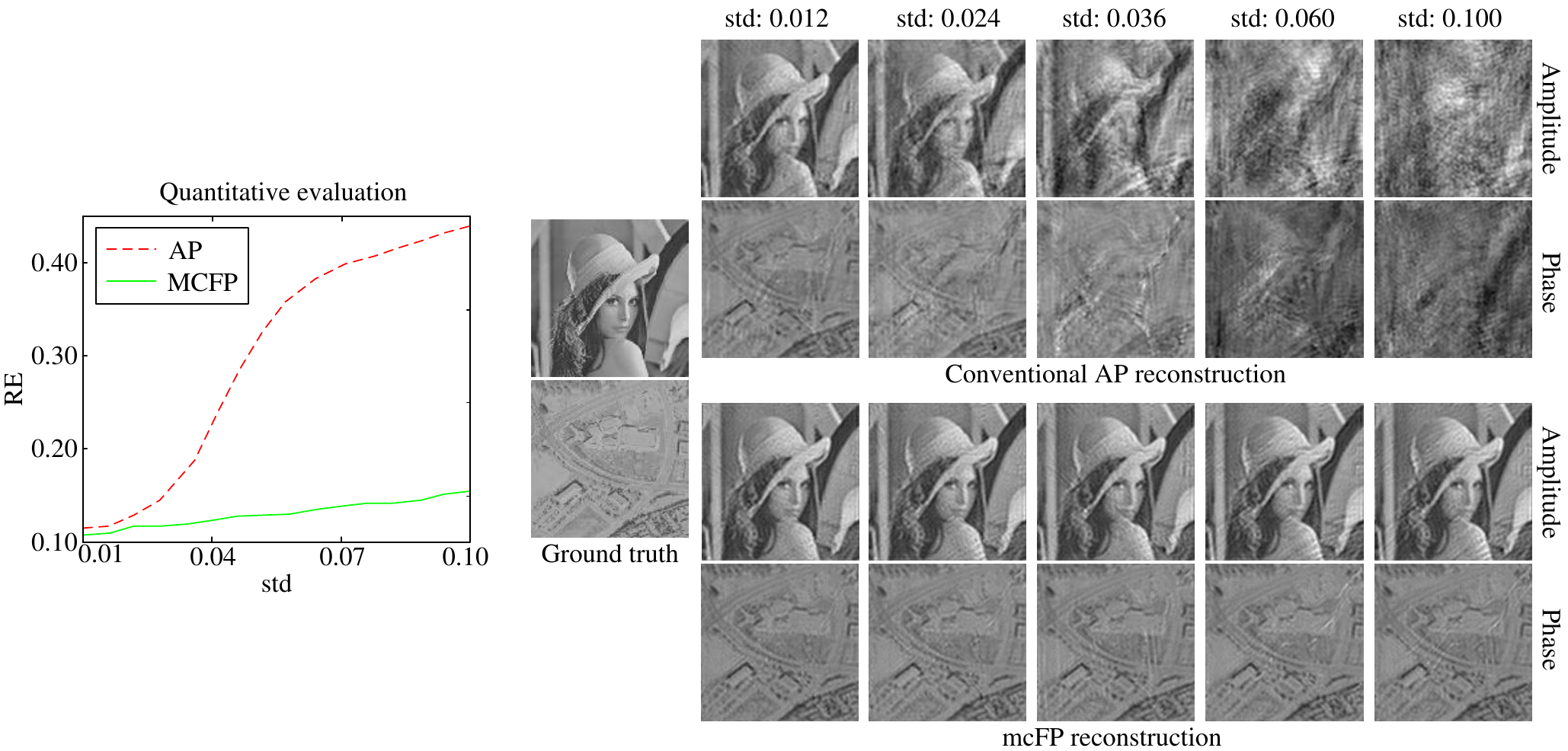}}
\caption{Reconstruction comparison between the proposed mcFP method and conventional AP algorithm, in the case with ideal circular sample motion. The quantitative evaluation is plotted on the left side, and the reconstructed amplitude and phase images by the two methods are shown on the right side.}
\label{fig:Fig_Simulation}
\end{figure}

First, we test the proposed mcFP method and conventional AP algorithm on simulated data in the case with ideal circular sample motion, which conforms to the assumption of mcFP in Eqs. (\ref{eqs:Formation_1}). Random Gaussian-distributed motion shift is added to the sample while capturing each LR image. For the AP algorithm, 100 iterations are enough to converge as proved in Ref. \cite{FPM_Nature}. We set 2 iterations for the proposed mcFP method which is experimentally validated and works well for successful reconstruction (note that in each iteration, we update each sub-spectrum for 10 times after the optimum motion shift guess is obtained). 
The reconstruction results are shown in Fig. \ref{fig:Fig_Simulation}, where the left sub-figure plots the quantitative evaluations, and the reconstructed images are shown on the right side. The standard deviation (std) of the motion shift is normalized by the scale of the system's FOV. From the results, we can see that conventional AP reconstruction degrades a lot as the sample motion becomes severe. Instead, mcFP works well to correct for the sample motion, and is barely affected even in the severe case (e.g., see the results when the std of motion shift is 10$\%$ of the FOV scale). This benefits from the additional motion recovery procedure and corresponding phase offset compensation, as well as the complete match between the image formation model (Eqs. (\ref{eqs:Formation_1})) and the reconstruction model (Eq. (\ref{eqs:Update})).

\begin{figure}[!t]
\centering
\centerline{\includegraphics[width=1.3\textwidth]{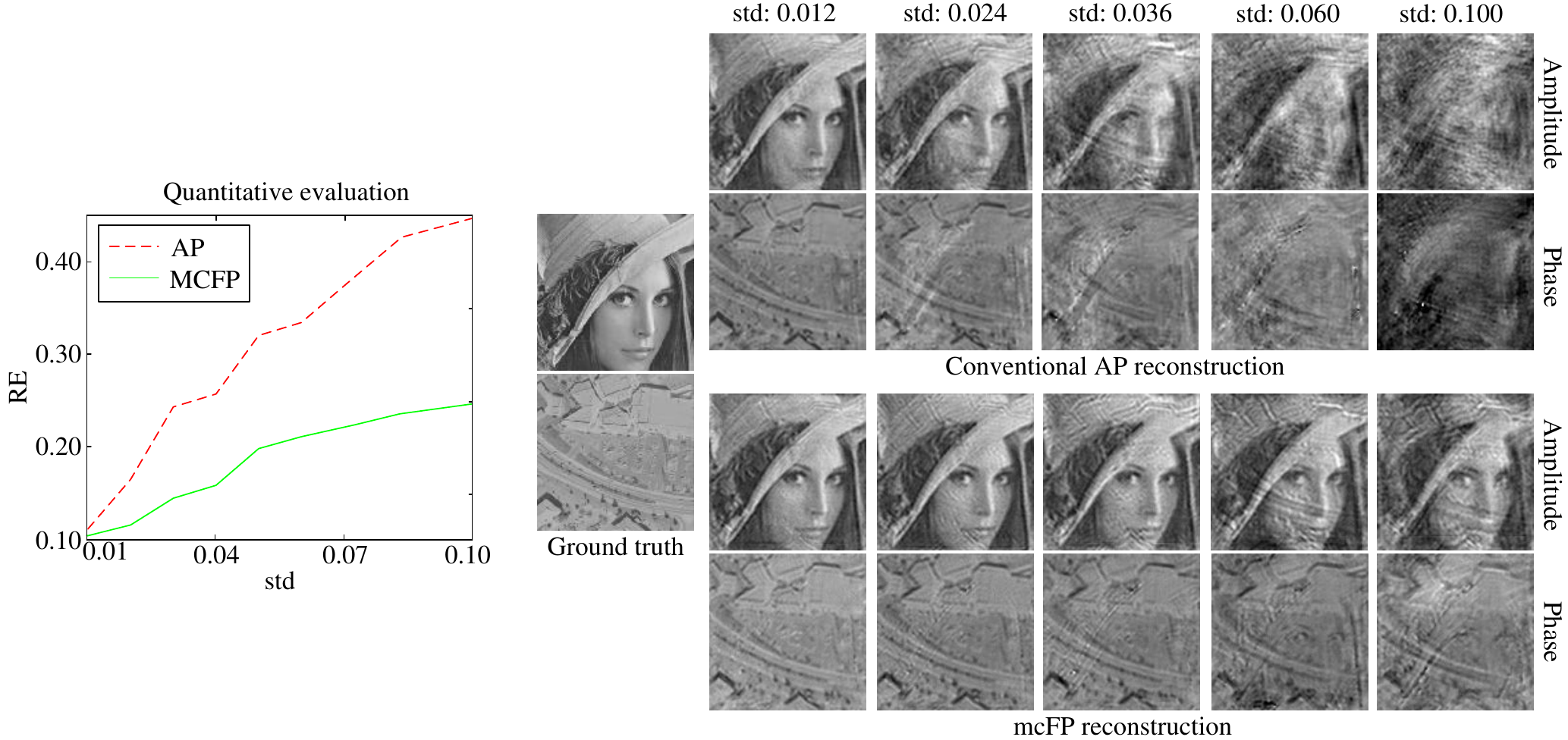}}
\caption{Reconstruction comparison between the proposed mcFP method and conventional AP algorithm, in the case with non-ideal circular sample motion (the sample region moving outside of the FOV is not the same as that moving inside).}
\label{fig:Fig_Simulation_NotCirc}
\end{figure}

However, the assumption of ideal circular sample motion is usually not valid for real applications, where the sample region moving outside of the FOV is usually not the same as that moving inside.
Next, we consider the real case. 
To simulate this, we set the FOV as the central region of the sample (central $256\times256$ pixels out of the entire $512\times512$ pixels), and simulate the sample motion by shifting the entire HR image. The corresponding reconstruction results by conventional AP and the proposed mcFP are shown in Fig. \ref{fig:Fig_Simulation_NotCirc}, from which we can see that mcFP still works well. Even in the case with severe sample motion (e.g. std = 0.1), though mcFP produces some unpleasant artifacts, it still outperforms conventional AP a lot. This validates the robustness and effectiveness of mcFP.

\subsection{Real experiment}

\begin{figure}[!t]
\centering
\centerline{\includegraphics[width=1.1\textwidth]{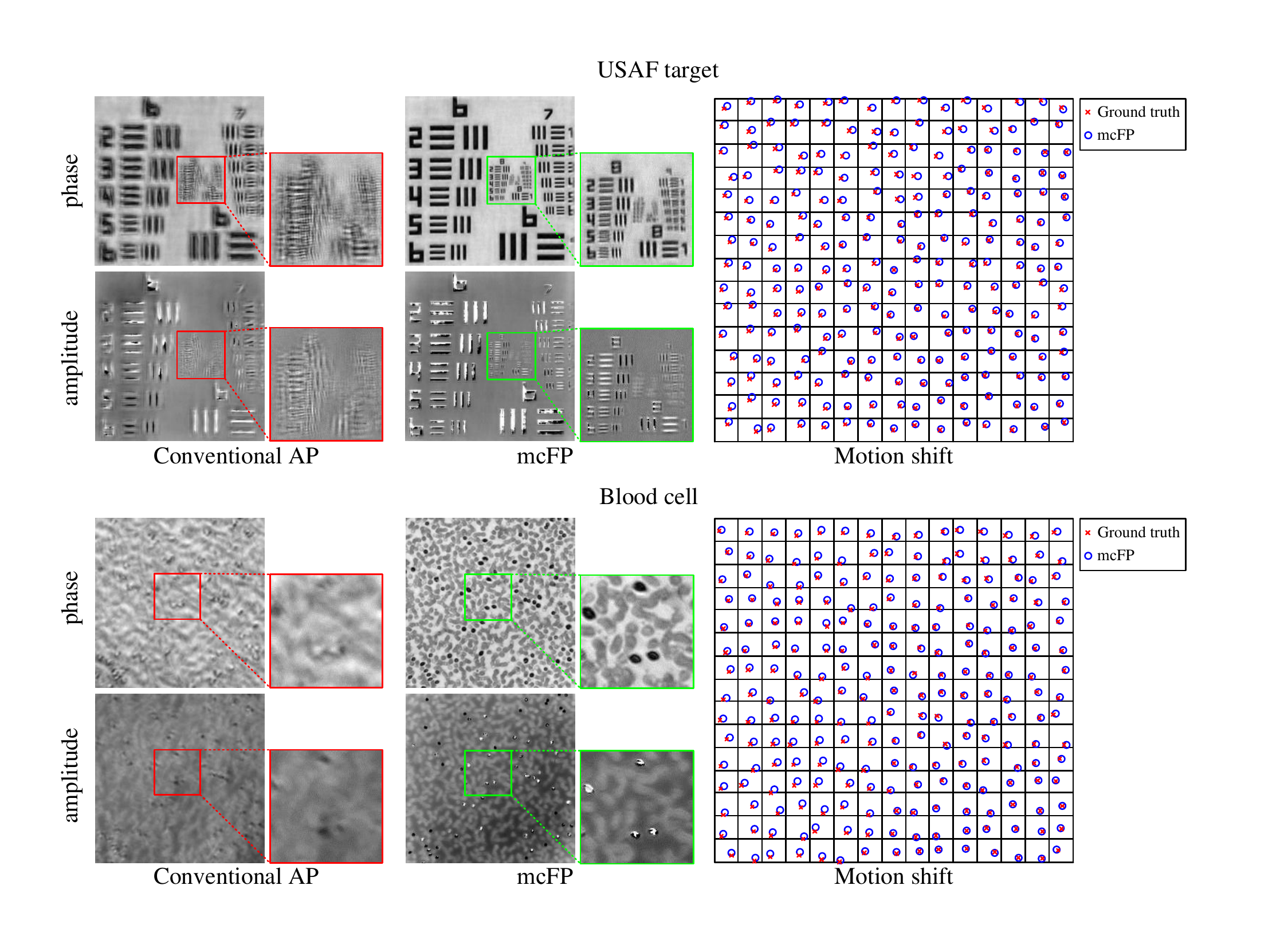}}
\vspace{-1mm}
\caption{Reconstruction results by conventional AP and the proposed mcFP on real captured datasets (USAF target and red blood cell) by an FPM setup. We also show the ground-truth motion shift and corresponding reconstruction of each measurement on the right, where each sub-square stands for the motion space of the sample when capturing corresponding LR image, and the boundary represents the maximum motion shift being 0.04 of the system's FOV scale.}
\label{fig:Fig_Real}
\end{figure}

To further validate the effectiveness of the proposed mcFP method, we test it on two real captured datasets including a USAF target and a red blood cell sample. 
Similar to the previous reported FPM setup \cite{FPM_Nature}, we used a regular optical microscope platform with a 2X, 0.1 NA Nikon objective lens in this experiment. A 15 $\times$ 15 LED array was used for sample illumination, and the incident wavelength is 632 nm. The distance between adjacent LED elements is 4 mm, and that between the LED array and the sample is 88 mm, corresponding to a maximum illumination NA of 0.45. In the acquisition process, we randomly translated the sample to different x-y positions and captured corresponding low-resolution images using a monochromatic CCD (GS3-U3-91S6M, Point Grey). The random x-y motions of the sample range from 0 to 10 microns, and they were treated as unknown parameters in the recovery process. The captured 225 low-resolution images were then used to recover the high-resolution sample image and the 225 x-y motion parameters utilizing the proposed mcFP scheme.

The reconstruction results are shown in Fig. \ref{fig:Fig_Real}. 
From the results we can see that AP can not work well and produces many unpleasant artifacts in the reconstructed images. mcFP obtains much better results than AP. For example, AP can only resolve the feature of group 6 on the USAF target, while mcFP can even resolve the group 9. The large resolving gap also exists in the results of the blood cell sample. Note that there are several phase discontinuity regions in the reconstructed phase maps of mcFP. This is because in these regions the reconstructed amplitude is close to zero, and the phase can be any value without affecting the final successful convergence. Besides, we also visualize the ground-truth motion shift and corresponding reconstruction of each captured LR image on the right side in the figure, where the precise match between them validates mcFP's ability to successfully recover unknown sample motion shift.
To conclude, mcFP can effectively correct for sample motion and reconstruct high-quality samples with much less artifacts, higher image contrast and more image details.

\section{Discussion}

In this paper, we propose a novel FP reconstruction method termed motion-corrected Fourier ptychography (mcFP), which adds an additional sample-motion recovery and phase-offset compensation procedure to conventional AP reconstruction, and adaptively updates the HR spatial spectrum from low spatial-frequency regions to high spatial-frequency regions.  
Results on both simulated data and real captured data show that the proposed mcFP method can successfully recover unknown sample shift and reconstruct high-quality HR amplitude and phase of the sample. It may find wide applications in related FP platforms such as endoscopy and transmission electron microscopy.

The reason that the proposed mcFP method works well for motion correction lies in two aspects. First, it utilizes the adaptive updating manner loop by loop from low spatial-frequency regions to high spatial-frequency regions. Since the overlap region of the spatial spectrum between the previous and the current loop has been updated in the previous loop, it offers an accurate guide for the precise motion shift guess of the current loop. Second, mcFP does not directly search for the solution to the phase aberration of each spatial frequency caused by sample motion, in which case a lot of unknown variables need to be determined. Instead, it searches for the precise sample motion shift in spatial space that owns only two unknown variables (namely $\Delta x$ and $\Delta y$). This largely narrows the variable space and guarantees successful recovery.

We note that the aberration caused by sample motion is different from that of pupil location error \cite{FPMAlgorithm_Laura, TPWFP}. For pupil location error which may be introduced by LED misalignment in conventional FPM setup, it produces the shift of the sample's spatial spectrum in Fourier space (the pupil plane), thus we obtain an intensity image corresponding to a different sub-region of the spatial spectrum. Differently, the sample motion we target to correct for in this paper is the shift of the sample in spatial space. It causes phase aberration of the sample's spatial spectrum instead of shift.

mcFP can be widely extended. First, the pupil function updating procedure of the EPRY-FPM algorithm \cite{PupilFunction} can be incorporated into mcFP to obtain corrected pupil function and better reconstruction. Second, the gradient descent method or the simulated annealing method\cite{luenberger1973introduction} can be applied to the motion shift search procedure (Eq. (\ref{eqs:Search})) in mcFP to accelerate the algorithm. 
Third, the current AP updating method (Eq. (\ref{eqs:Replace}) and Eq. (\ref{eqs:Update})) can be replaced with more robust methods such as WFP \cite{FPMAlgorithm_Bian} and TPWFP \cite{TPWFP}, which would improve the reconstruction quality, especially in the case with severe sample motion. Fourth, in the above we assume that the sample is fixed during the exposure time of each individual acquisition, because the sample motion is not continuous for the above mentioned systems including endoscope and TEM. Besides, the exposure time can be very short \cite{LaserFPM}. However, there still exists possibility that the sample moves during the exposure time, which causes motion blur in the captured image. In this case, we can add an additional deblurring procedure to mcFP by either the blind deconvolution methods \cite{fish1995blind, shan2008high, cho2009fast} or other methods \cite{hansen2006deblurring}. Thus we can find the latent blur-free image, and then perform the above spectrum updating for reconstruction. This would help when we apply FP for imaging live samples.


\section*{Acknowledgements}

This work was supported by the National Natural Science Foundation of China (Nos. 61120106003 and 61327902). G. Zheng acknowledges the support of the National Institutes of Health (NIH 1R21EB022378-01).

\end{document}